\documentclass[fleqn,twoside]{article}
\usepackage{espcrc2}

\usepackage{graphicx}

\usepackage[figuresright]{rotating}

\newcommand{\AmS}{{\protect\the\textfont2
  A\kern-.1667em\lower.5ex\hbox{M}\kern-.125emS}}

\title{State transitions in the 2001/2002 outburst of XTE J1650--500}

\author{S. Rossi\address[AOB]{INAF-Astronomical Observatory of Brera-Merate,\\
        Via E.Bianchi 46, 23807 Merate (LC), Italy},
        J. Homan\addressmark[AOB],
        J.M. Miller\address[HSCA]{Harvard-Smithsonian Center for Astrophysics,\\
        60 Garden Street,Cambridge, MA 02138, U.S.A.},
         T. Belloni\addressmark[AOB]}
\begin{document}

\begin{abstract}

We present a study of the X-ray transient and black hole candidate
XTE J1650--500 during its 2001/2002 outburst. The source made two
state transitions between the hard and soft states, at luminosity
levels that differed by a factor of $\sim$5--10. The first
transition, between hard and soft, lasted for $\sim$30 days and
showed two parts; one part in which the spectral properties evolve
smoothly away from the hard state and another that we identify as
the 'steep power law state'. The two parts showed different
behavior of the Fe K emission line and QPO frequencies. The second
transition, from soft to  hard, lasted only $\sim$15 days and
showed no evidence of the presence of the 'steep power law state'.
Comparing observations from the early rise and the decay of the
outburst, we conclude that the source can be in the hard state in
a range of more $10^4$ in luminosity. We briefly discuss the state
transitions in the framework of a two-flow model. \vspace{1pc}
\end{abstract}

\maketitle

\section{INTRODUCTION}

Black Hole Candidate (BHC) X-ray transients often show correlated
spectral and variability properties during their outbursts. Several
distinct states of spectral and variability behavior have been
recognized (\cite{tale1995,va1995a,mcre2003}). Until a few years ago,
it was thought that the states of a source and transitions between
them were determined and driven by (variations in) the mass accretion
rate ($\dot{m}$). Recent observations
(\cite{ruleva1999,howiva2001,wido2001,retrbo2000,trbopr2001,maco2003,rohomi2003})
have challenged this idea; these suggest that in addition to
$\dot{m}$ there is another parameter responsible for the state of a
system.

Here we present a study of the state transitions observed with the
{\it Rossi High Energy Timing Explorer} ({\it RXTE}) during the
2001/2002 outburst of the BHC X-ray transient XTE J1650--500.

\begin{figure}[ht] \resizebox{\hsize}{!}{\includegraphics{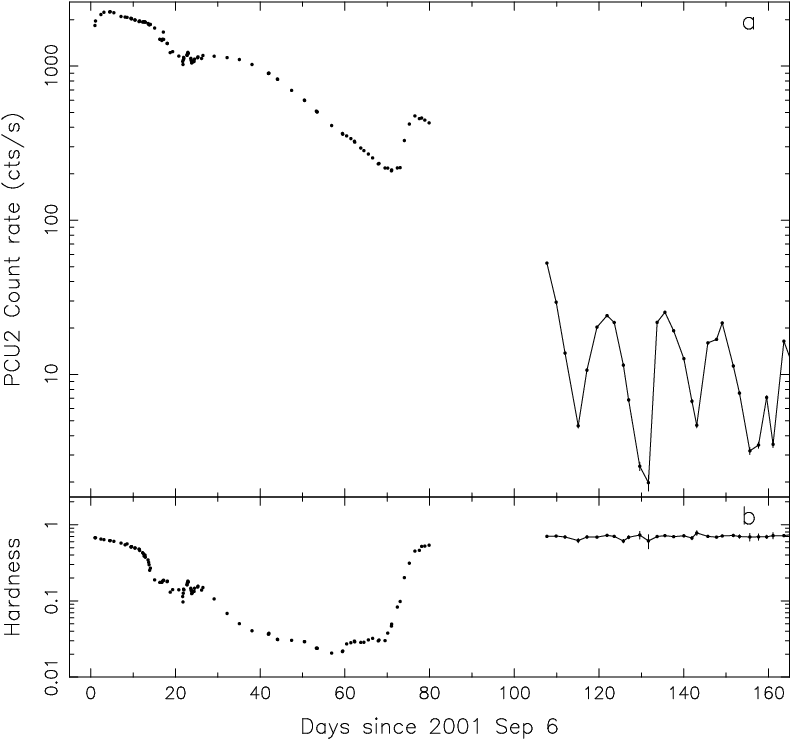}}
\caption{PCU2 light curve (a) and hardness ratio (b) of the first 185
days of outburst. Count rate is in the 2.5-25 keV band and the
hardness is the ratio of counts in the 4.6-7.9 keV and 2.5-4.6 keV
bands. } \label{fig:lc_hr} \end{figure}

\begin{figure}[ht] \resizebox{\hsize}{!}{\includegraphics{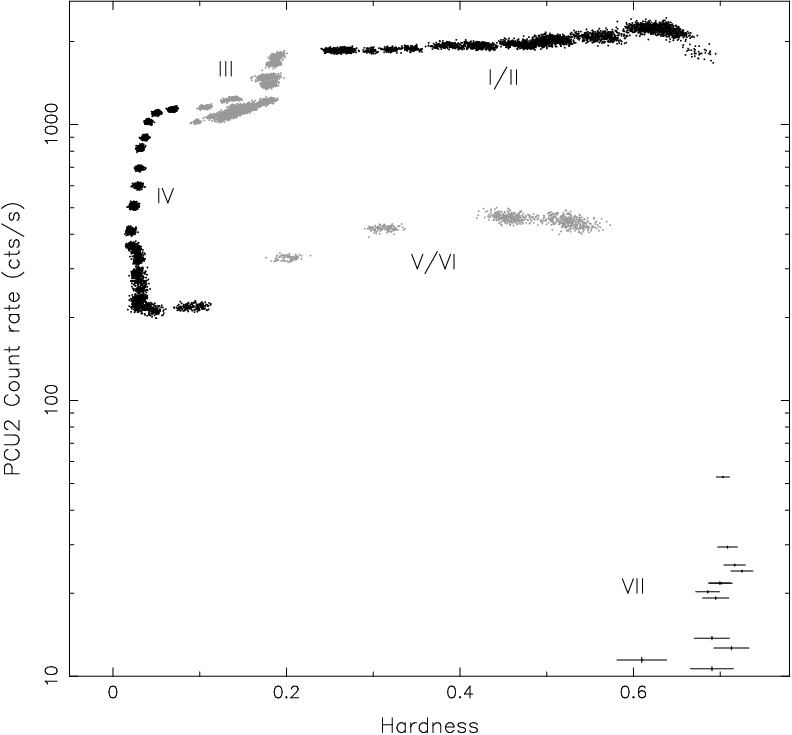}}
\caption{A hardness-intensity diagram showing distinct groups in
the spectral evolution. XTE J1650--500 described a
counter-clockwise track, starting in the upper-right corner (I/II)
and ending in the lower-right one (VII).} \label{fig:hid}
\end{figure}

\begin{figure}[!ht]
\resizebox{\hsize}{!}{\includegraphics{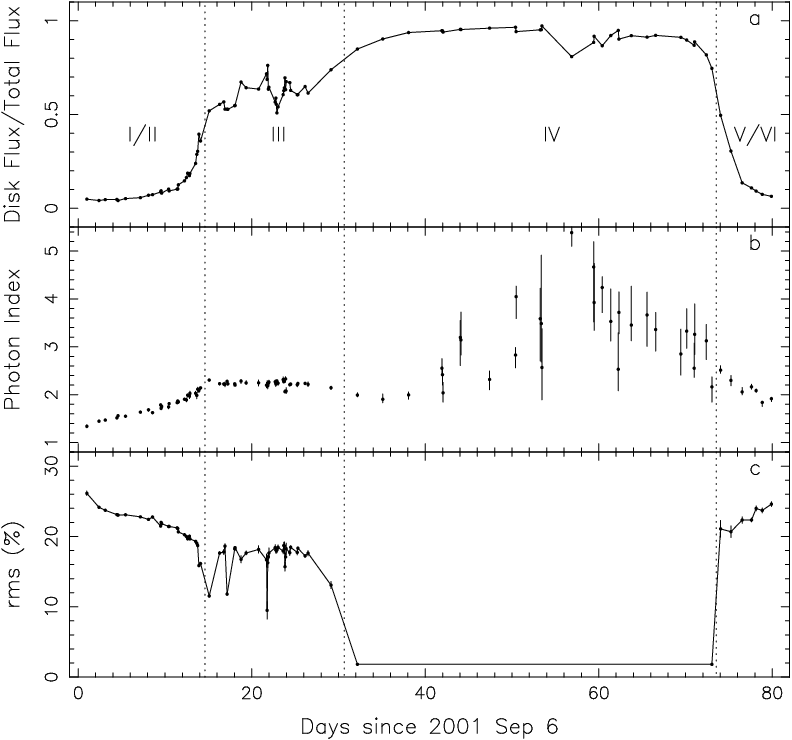}}
\caption{Evolution of spectral and variability parameters from
Sep.6 to Nov. 24: 2-100 keV unabsorbed disk flux normalized to the
total flux (a); photon index of power-law component (b);
variability (c). The vertical lines mark the transitions between
the first four groups} \label{fig:states}
\end{figure}

\section{OBSERVATIONS AND RESULTS}

\subsection{Data analysis}

We used 176 pointed {\it RXTE} observations for our study of XTE
J1650--500, covering the period between 2001 September 6 and 2002
June 21. Especially the beginning of the outburst was extremely well
sampled, giving us the opportunity to study the hard$\rightarrow$soft
transition in great detail. For our spectral analysis, we combined
PCA and HEXTE data to obtain 3--150 keV energy spectra. For our
timing analysis, we performed Fast Fourier Transforms (FFTs) of the
high time resolution data (6--32 keV), resulting in 1/128--1024 Hz
power density spectra. All energy spectra were fitted with a
combination of a (cut-off) power law, a disk black body, a smeared
edge, and a relativistic Fe K emission line from the accretion disk;
our power spectra were fitted with a model consisting of several
Lorentzians.

\subsection{Overall outburst behavior}

The 2.5--25 keV PCA light curve and the 4.5--7.9 keV/2.5--4.6 keV
hardness ratio are shown in Figure \ref{fig:lc_hr}. Already before
the peak in the count rate the spectrum was softening. This trend
continued for the early part of the decay until day 16, when both
the count rate and hardness showed erratic changes. After day 33
the smooth softening observed before day 16 continued, which was
accompanied by a further decrease in the count rate. On day 75 the
spectrum showed a sudden hardening and an increase in the count
rate \cite{katoro2003}. After a 30 day gap in our observations
(due to solar constraints) the source had a more or less constant
spectral hardness similar to that of the first observation, while
the count rate showed long term oscillations on top of a slow
decay \cite{tokaco2003,tokaka2003}. The complex relation between
count rate and spectral hardness can also be seen in the
hardness-intensity diagram (HID; Fig.~\ref{fig:hid}), which shows
several distinct branches. Starting in the upper-right corner, the
source moved through the HID counter-clockwise. Although it
returned to a similar hardness as in the beginning of the
outburst, it did so at a count rate that was a factor of 30 lower.
Following \cite{howiva2001} we defined several groups, based on
their position in the HID, the spectral and variability properties
of which will be the subject of the next section. An analysis of
RXTE/ASM data strongly suggests that during the rise of the
outburst the source moved along a path of constant hardness
($\sim0.7$) and increasing count rate to the right-hand side of
group I/II.

\begin{figure}[ht]
\resizebox{\hsize}{!}{\includegraphics{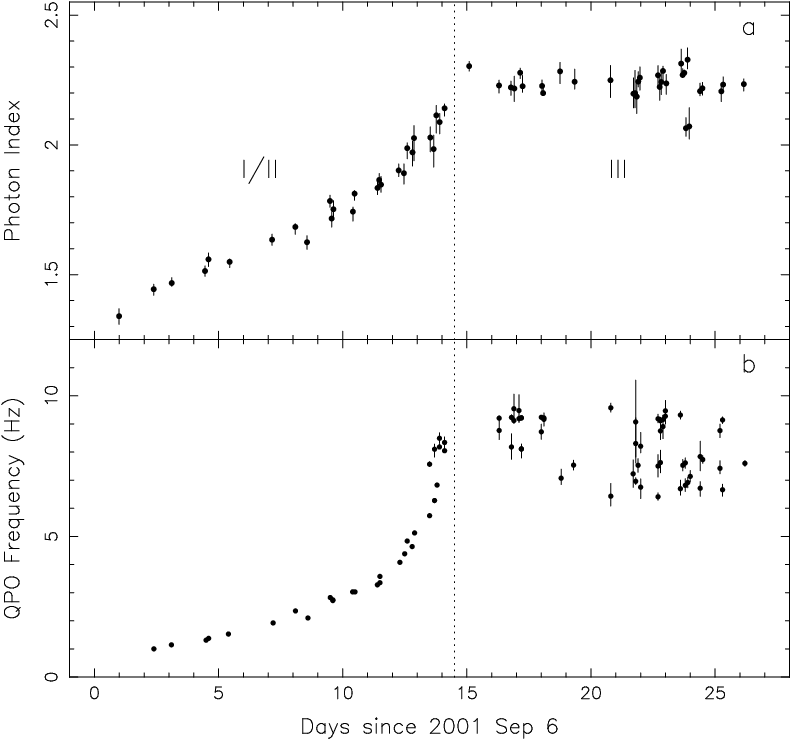}}
\caption{Evolution of power law index (a) and the low-frequency
QPO (b) from Sep. 6 to Oct. 3. The vertical line marks the
transition from group I/II to group III.} \label{fig:pli_qpo}
\end{figure}

\subsection{Correlated spectral and variability properties}

\begin{figure}[ht]
\resizebox{\hsize}{!}{\includegraphics{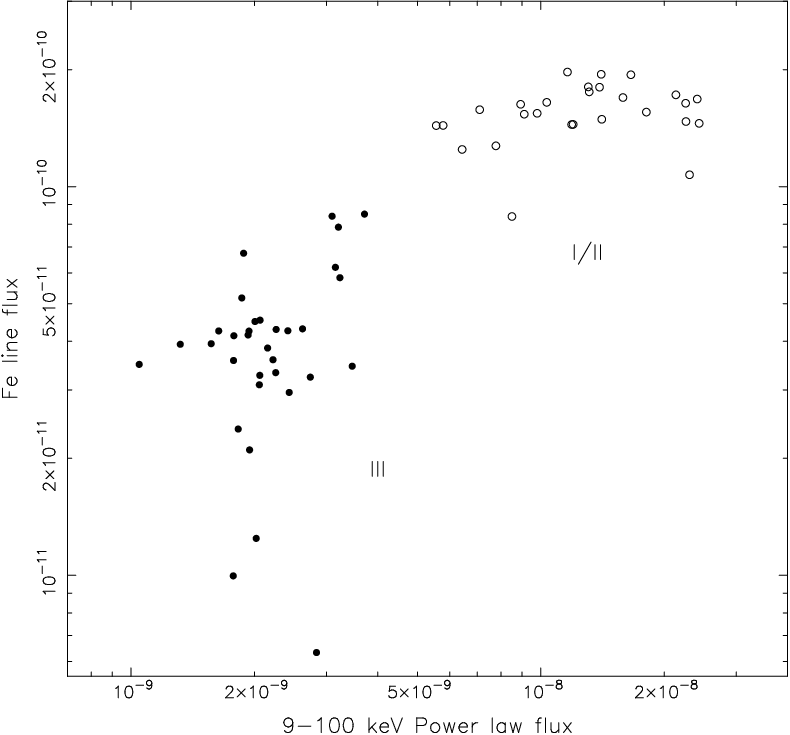}} \caption{The
iron line flux as a function of 9--100 keV power law flux for groups
I/II and III.} \label{fig:line} \end{figure}

Like in other BHC transients, the energy and power spectra evolved in
a correlated way during the outburst. In XTE J1650--500 the types of
correlated behavior changes from one group to the other. This is
clear from Figure \ref{fig:states} in which we show the fractional
contribution of the disk component ($f_d$) in the 2--100 keV range
(\ref{fig:states}a), the index ($\Gamma$) of the power-law component
(\ref{fig:states}b) and the strength ($r$) of the 0.01--100 Hz
variability (\ref{fig:states}c). The three vertical lines mark the
transitions between the first four groups, each of which shows
distinct behavior - data from after the gap are not shown. At the
beginning of the outburst the source showed properties that are
typical for the hard state ($f_d<$0.05, $\Gamma$=1.35, and $r$=26\%
rms). In group I/II these parameters smoothly changed to
$f_d\sim$0.4, $\Gamma$=2.1, and $r$=16\% rms, which is typical for
the very high state. At the time of the transition to group III the
2--100 keV flux had decreased from  $3.4\,10^{-8}$
erg~cm$^{-2}$~s$^{-1}$, at the peak of the outburst (third day), to
$1.8\,10^{-8}$ erg~cm$^{-2}$~s$^{-1}$. In group III  $f_d$ and $r$
changed erratically, whereas $\Gamma$ was remarkably constant
($\sim$2.25). Although the properties of group III classify it as
very high state, its behavior is quite different from that in group
I/II as is evident from Figs.~\ref{fig:pli_qpo} and \ref{fig:line}
(more on those later). In group IV the spectrum was dominated by the
disk component ($f_d>0.85$) and variability was very weak ($r<2\%$
rms), typical for the soft state. Group V/VI was characterized by the
sudden decrease in $f_d$ ($<$0.3) and the return of strong
variability ($r\sim$22\%); $\Gamma$ smoothly decreased to 1.8 --
these properties were very similar to observations in group I/II,
albeit at a flux that was a factor of $\sim$5--10 lower. After the
$\sim$30 day gap in our observations (group VII) $\Gamma$ was similar
to that in the first observation and variability was strong
($r\sim$35\%), typical for the hard state. For a discussion of the
remarkable $\sim$14 day oscillations in this phase of the outburst we
refer to \cite{tokaco2003,tokaka2003}.

\section{DISCUSSION}

During its 2001/2002 outburst XTE J1650--500 showed several types
of behavior, which could be classified either as distinct states
(groups III and IV, VII) or as transitions between states (groups
I/II and V/VI). Only during the transition from the hard to the
soft state an extended interval (group III) of nearly constant
$\Gamma$ was observed. Interestingly, this is the same part of the
outburst in which narrow high frequency QPOs were found by
\cite{hoklro2003}. This is most likely the 'steep power-law'
state, as defined by \cite{mcre2003}. Figure \ref{fig:pli_qpo}
shows that not only $\Gamma$ saturates in this state, but the QPO
frequency as well, although it is not completely clear whether the
QPOs in groups I/II and III are of the same type. The behavior of
the Iron line flux as a function of power law flux
(Fig.~\ref{fig:line}) also reveals different patterns for groups
I/II and III, suggestive of a change in the properties of the hard
component and/or ionizations state of the disk \cite{rohomi2003}.
Additional changes are observed in the high energy cutoff and the
coherence and time lag properties. The fact that this state is not
observed during the soft$\rightarrow$hard transition suggests that
this transition does not necessarily display the reverse behavior
seen during the hard$\rightarrow$soft transition as was suggested
by \cite{howiva2001}.

The start of the hard$\rightarrow$soft transition occurred at an
2--100 keV luminosity that was a factor of $\sim$10 higher than at
the start of the soft$\rightarrow$hard transition. This hysteresis
effect, which was already recognized  as such by \cite{mikiha1995}
in other transients, has been attributed by several authors to the
presence of two flows in these X-ray binary systems
\cite{maco2003,va2001,smhesw2002}; one which responds rapidly to
changes in the mass accretion rate (a coronal type flow) and one
that responds much slower (a disk flow). The fact that almost the
entire rise likely took place in the hard state, suggests that the
fast flow couples directly to the medium responsible for the hard
flux. Although the hard states in black hole transients are
observed over a range of more than $10^4-10^6$ in luminosity, they
only seem to be stable at the low luminosity end. That they are
still observed at high luminosities, for short periods, is likely
the result of the slow response of the disk flow. Whether the
eventual decrease in the hard component is due to the
disappearance of the fast flow, or  to another mechanism, possibly
involving cooling by the soft disk flux, is not clear. What is
clear, though, is that like in many other systems \cite{ma2003},
the hard component in XTE J1650--500 starts to dominate again at
luminosities of less than $\sim$10\% of the Eddington luminosity.

\end{document}